\documentclass[12pt]{article} \topmargin= 0.3cm \textwidth 16.5cm
=0cm \headsep =0cm
\usepackage {latexsym}
\usepackage {graphicx}

\newcommand{\p}{\partial}
\newcommand{\pr}{\prime}
\newcommand{\pp}{{\prime \prime}}
\newcommand{\be}{\begin{equation}}
\newcommand{\ee}{\end{equation}}

\begin{document}

\title{Graviton localization and Newton's law for brane models with a non-minimally coupled bulk scalar field}
\author{K. Farakos \footnote{kfarakos@central.ntua.gr}, G. Koutsoumbas \footnote{kutsubas@central.ntua.gr} and P. Pasipoularides  \footnote{paul@central.ntua.gr} \\
       Department of Physics, National Technical University of
       Athens \\ Zografou Campus, 157 80 Athens, Greece}
\date{ }
       \maketitle

\begin{abstract}
Brane world models with a non-minimally coupled bulk scalar field
have been studied recently. In this paper we consider metric
fluctuations around an arbitrary gravity-scalar background
solution, and we show that the corresponding spectrum includes a
localized zero mode which strongly depends on the profile of the
background scalar field. For a special class of solutions, with a
warp factor of the RS form, we solve the linearized Einstein
equations, for a point-like mass source on the brane, by using the
brane bending formalism. We see that general relativity on the
brane is recovered only if we impose restrictions on the parameter
space of the models under consideration.
\end{abstract}

\section{Introduction}

The old Kaluza-Klein idea of enlarging the space-time manifold to
extra dimensions has been put on a new basis recently, in an
attempt to solve the hierarchy problem between the electroweak
scale and the Planck scale. The new point is that there is no need
to consider extra dimensions of the order of the Planck length:
one may consider large compact extra dimensions of the order of a
millimeter \cite{ADD} or even a non-compact extra dimension
\cite{Ran}. An interesting feature of these models, is that they
have testable predictions in the near future high energy
experiments.

The localization problem is a major issue in these so-called brane
world models \cite{Akama,Shap}. The graviton propagates in all
dimensions (also in the bulk), since it is the dynamics of
spacetime itself. The SM particles are localized in some way on
the $4D$ submanifold (Brane), which corresponds to the universe we
are living in. Several localization mechanisms have been proposed;
see for example \cite{Shif,Rub} and references therein. For
lattice simulations on the same subject see \cite{lat}.

In \cite{ADD} one introduces $n$ flat extra compact spatial
dimensions with large volume. This brane world model is known as
ADD model (Arkani-Hamed, Dimopoulos and Dvali). An important
feature of a wide class of brane world models, including the ADD
model, is that they predict deviations from the 4D Newton law at
submillimeter distances.

The simplest brane models with a warped extra dimension have been
introduced in \cite{Ran}. In the first version, which is known as
the first Randall-Sundrum model, we have an orbifolded extra
dimension of radius $r_c$. Two branes are fitted to the fixed
points of the orbifold, $z=0$ and $z_c=\pi r_c$ with tensions
$\sigma$ and $-\sigma$ respectively; it is assumed that the
particles of the standard model are trapped on the negative
tension brane. The second version of the model is constructed if
we send the negative tension brane to infinity ($r_c\rightarrow
+\infty$) and assume that the ordinary matter lives on the
positive tension brane.

In both versions of the Randall-Sundrum model the $4D$ graviton is
obtained by considering small gravitational fluctuations
$\bar{h}_{\mu\nu}$ around the classical solution of the model. The
spectrum of gravitational fluctuations $\bar{h}_{\mu\nu}$ consists
of a zero mode, which gives rise to Newton's law on the brane,
plus a continuum of positive energy states with no gap, giving
small corrections to the $4D$ Newtonian potential. For a
derivation of the equation obeyed by the metric fluctuations see
for example \cite{wolfe}.

The proper calculation of the Newtonian potential on the brane
plus the correction terms has been detailed by J. Garriga and T.
Tanaka \cite{GT} and  by S.B. Giddings et al \cite{RN}. These
works use the so-called bent brane formalism. The idea is that if
one puts a matter source on the brane, there is a gauge choice for
which the equations for the metric perturbations decouple;
however, in this gauge the brane is not located at $z=0$ but it
appears bent around the position of the matter source. In
\cite{GT} it is pointed out that the role of the bending is
essential in reproducing the $4D$ graviton propagator structure
out of the $5D$ one, as it exactly compensates for the effects of
extra polarization. Calculations of the Newton potential based on
different philosophies can be found in \cite{ADM,CR,VV,SV}. For a
recent review on the topic the reader may consult \cite{BKM} and
references therein.

There are various generalizations of the ideas of standard
Randall-Sundrum scenario, such as models with more than five
dimensions, models with topological defects toward the extra
dimensions, multi-brane models and models with higher order
curvature corrections (i.e. Gauss-Bonnet gravity). Details are to
be found in, for example, \cite{Rub,Lon,AH} and references
therein.

Brane world models with a non-minimally coupled bulk scalar field,
via an interaction term of the form $-\frac{1}{2}\xi R \phi^2$,
where $\xi$ is a dimensionless coupling, have been studied
recently. Static solutions of these models have been examined
numerically in \cite{FP,KF,KFR}, for the simplest case of a scalar
field potential $V(\phi)=\lambda \phi^4$, while in \cite{KFGB} the
same model has been examined in the presence of Gauss-Bonnet
gravity. In \cite{Tam} analytical solutions have been obtained by
choosing appropriately the potential for the scalar field.
Cosmological implications of these models are described in
\cite{Tamcos}.

A crucial question is, whether standard four-dimensional gravity can
be recovered on the brane in models with a non-minimally coupled
scalar field. In order to answer this question we consider metric
fluctuations around an arbitrary gravity-scalar background solution
of these models. We see that the corresponding spectrum does include
a localized zero mode, which is necessary for Newton's law to hold
on the brane. For a special class of solutions, with a warp factor
of the RS form, we solve the linearized Einstein equations in the
case where a point-like mass source sits on the brane, using the
bent brane formalism. We see that standard four-dimensional gravity
on the brane is recovered only if we impose serious restrictions on
the parameters of these models.

In section 2 of this work we define the model; in section 3 we
give the first order (linearized) version of the Ricci and the
energy-momentum tensors, concluding with an equation for the
gravitational fluctuation in a source free space-time; in section
4 we comment on the spectrum of the metric fluctuations. In
section 5 we present the bent brane formalism for our case, we
calculate the gravitational perturbations and comment on the
results. Section 6 summarizes our conclusions. Finally, in an appendix
we give a short discussion on tensor fluctuations in the Einstein frame.

\section{The Model}

We consider the action of five-dimensional gravity with a
non-minimally coupled bulk scalar field:
\begin{equation}
 S=\int d^{5}x \;\sqrt{|g|}\left(
F(\phi)R-\frac{\sqrt{|g^{(brane)}|}}{\sqrt{|g|}}\sigma(\phi)\delta(z)
-\frac{1}{2}g^{\mu\nu} \nabla_{\mu}\phi
\nabla_{\nu}\phi-V(\phi)\right),
\end{equation} \label{actio1}
where
\begin{equation}
F(\phi)=\frac{1}{2 k}(1-k \xi \phi^2 ), \quad k=8 \pi G_{5}.
\end{equation}
$G_{5}$ is the five-dimensional Newton's constant, and
$\sigma(\phi)$ is a $\phi$-dependent brane tension. We have
assumed that $\mu, \nu=0,1,2,3,5$, and $d^{5}x=d^{4}x dz$ where
$z$ describes the extra dimension. Note that in the above action
we did not add an explicit cosmological constant term; if it is
present, it may be included in the scalar field potential
$V(\phi).$

The Einstein equations with the non-minimally coupled bulk scalar
field are:
\begin{equation}
F(\phi) G_{\mu\nu}-\nabla_{\mu}\nabla_{\nu}F(\phi)+ g_{\mu\nu}\Box
F(\phi)+\frac{1}{2}\sigma(\phi)
\delta(z)\frac{\sqrt{|g^{(brane)}|}}{\sqrt{|g|}}
g_{ij}\delta_{\mu}^{i}\delta_{\nu}^{j}=\frac{1}{2}T^{(\phi)}_{\mu\nu}
\label{Einstein1}
\end{equation}
where $i,j=0,1,2,3$, $G_{\mu\nu}=R_{\mu\nu}-\frac{1}{2}g_{\mu\nu}R$
, and $T^{(\phi)}_{\mu\nu}$ is the energy momentum tensor for the
scalar field
\begin{equation}
T^{(\phi)}_{\mu\nu}=\nabla_{\mu}\phi\nabla_{\nu}\phi-g_{\mu\nu}
\left[\frac{1}{2}g^{\rho\sigma}\nabla_{\rho}\phi\nabla_{\sigma}\phi+V(\phi)\right].
\end{equation}

The equation of motion for the scalar field reads:
\begin{equation}
\Box \phi+F'(\phi)R-V'(\phi)-\sigma'(\phi)\delta(z)=0,
\label{scalar1}
\end{equation}
where $F'(\phi) \equiv \frac{dF(\phi)}{d\phi};$ similar
definitions are understood for $V'(\phi)$ and $\sigma'(\phi).$

The equations of motion (\ref{Einstein1}) and (\ref{scalar1})
possess static solutions of the form:
\begin{equation}
ds^2=e^{2 A(z)}\eta_{ij}dx^{i}dx^{j}+dz^2, \quad \phi=\phi(z)
\label{background}
\end{equation}
which exhibits four-dimensional Poincar\`{e} symmetry. The sign
convention for the Minkowski metric is $\eta_{ij}=diag(-1,1,1,1)$.

Static solutions of the form of equation (6) have been studied
both numerically \cite{KF} and analytically \cite{Tam}.

\section{Linearized equations}

In this section we determine the linearized equations for the
metric fluctuations $\bar{h}_{\mu\nu}$, in the case of the brane
model with the non-minimally coupled scalar field introduced in
section 2.

It is convenient to consider a Gaussian normal coordinate system
$\bar{x}^{i},\bar{z}$, where the brane is located by definition at
$\bar{z}=0$, and the fluctuations $\bar{h}_{\mu\nu}$, around the
brane background solution, satisfy the conditions
$\bar{h}_{\mu5}=\bar{h}_{55}=0$. Note that we do not consider
scalar field fluctuations in the sequel; we will explain at the
end of this section that this is consistent.

The perturbed metric in this coordinate frame reads:
\begin{equation}
ds^2=e^{2
A(\bar{z})}\left(\eta_{ij}+\bar{h}_{ij}\right)d\bar{x}^{i}d\bar{x}^{j}+d\bar{z}^2
\end{equation}

The Ricci tensor may be expanded to read:
\begin{equation}
R_{\mu\nu}=R^{(0)}_{\mu\nu}+R^{(1)}_{\mu\nu}+..., \label{expr}
\end{equation}
where the zero order term is:
\begin{equation}
R^{(0)}_{ij}=- e^{2 A}(A''+4A'^2) \eta_{ij}, \quad
R^{(0)}_{55}=-4(A''+A'^2), \quad R^{(0)}_{5i}=0,
\end{equation}
while the first order term is:
\begin{eqnarray} R^{(1)}_{ij}=&-&e^{2A}(\frac{1}{2}\partial^2_z+2A'\partial_{z}
+A''+4A'^2)\bar{h}_{ij}-\frac{1}{2}\Box^{(4)}
\bar{h}_{ij}-\frac{1}{2}\eta_{ij}e^{2A}A'\partial_z(\eta^{kl}\bar{h}_{kl})\nonumber
\\&&-\frac{1}{2}\eta^{kl}(\partial_i\partial_j h_{kl}-\partial_i\partial_k
h_{jl}-\partial_j\partial_k h_{il}),\\&&
R^{(1)}_{55}=-\frac{1}{2}(\partial^2_z+2A'\partial_{z})\eta^{kl}\bar{h}_{kl},
\quad R^{(1)}_{i5}=\frac{1}{2}\eta^{kl}\partial_z(\partial_k
\bar{h}_{il}-\partial_i \bar{h}_{kl}),
\end{eqnarray}
where we have used the 4D d' Alembertian operator: $\Box^{(4)}
\equiv \eta^{ij}\partial_i\partial_j.$

In this paper we will use the alternative form of the Einstein
equations, reading:
\begin{equation}
2 F(\phi) R_{\mu\nu}=\tilde{t}_{\mu\nu}, \label{Einstein2}
\end{equation}
where $\tilde{t}_{\mu\nu}=t_{\mu\nu}-\frac{1}{3}g_{\mu\nu}t$,
$t=t^{\mu}_{\mu}=g^{\mu\nu}t_{\mu\nu}$, and $t_{\mu\nu}$ is
defined as
\begin{equation}
t_{\mu\nu} \equiv T^{(\phi)}_{\mu\nu}+2
\nabla_{\mu}\nabla_{\nu}F(\phi)-2 g_{\mu\nu}\Box
F(\phi)-\sigma(\phi)
\delta(z)\frac{\sqrt{|g^{(brane)}|}}{\sqrt{|g|}}
g_{ij}\delta_{\mu}^{i}\delta_{\nu}^{j}.
\end{equation}
The tensor $\tilde{t}_{\mu\nu}$ may also be expanded as:
\begin{equation}
\tilde{t}_{\mu\nu}=\tilde{t}^{(0)}_{\mu\nu}+\tilde{t}^{(1)}_{\mu\nu}+...,
\label{perscalar}
\end{equation}
with the zero order term:
\begin{equation}
\tilde{t}^{(0)}_{ij}=\frac{2}{3}\eta_{ij}e^{2A}\left((\phi')^{2}F''(\phi)+7
\phi'A'F'(\phi)+\phi''F'(\phi)+V(\phi)\right)-\frac{1}{3}\eta_{ij}
\sigma(\phi) \delta(z)
\end{equation}
\begin{equation}
\tilde{t}^{(0)}_{55}=(\phi')^{2}+\frac{8}{3}(\phi')^{2}F''(\phi)+\frac{8}{3}\phi'A'F'(\phi)+\frac{8}{3}\phi''F'(\phi)+\frac{2}{3}V(\phi),
\quad \tilde{t}^{(0)}_{5i}=0,
\end{equation}
and the first order term:
\begin{eqnarray}
\tilde{t}^{(1)}_{ij}&=&\frac{2}{3}e^{2A}\left((\phi')^{2}F''(\phi)+7
\phi'A'F'(\phi)+\phi''F'(\phi)+V(\phi)\right)\bar{h}_{ij}-\frac{1}{3}
\sigma(\phi)\delta(z)\bar{h}_{ij}\nonumber\\
&+&\frac{1}{3}\eta_{ij}e^{2A}F'(\phi)\phi'\partial_{z}(\eta^{kl}\bar{h}_{kl})
+e^{2A}F'(\phi)\phi'\partial_{z}\bar{h}_{ij},
\end{eqnarray}
\begin{eqnarray}
\tilde{t}^{(1)}_{55}=\frac{1}{3}e^{2A}F'(\phi)\phi'\partial_{z}(\eta^{kl}\bar{h}_{kl}),
\quad \tilde{t}^{(1)}_{5i}=0.
\end{eqnarray}

If we use equations (\ref{expr}), (\ref{Einstein2}) and
(\ref{perscalar}) we obtain the zero and first order Einstein
equations:
\begin{eqnarray}
&&2 F(\phi) R^{(0)}_{\mu\nu}= \tilde{t}^{(0)}_{\mu\nu},\\&& 2
F(\phi) R^{(1)}_{\mu\nu}= \tilde{t}^{(1)}_{\mu\nu}. \label{Efirst}
\end{eqnarray}

The background solution for the metric, equation
(\ref{background}), satisfies the zero order Einstein equations:
\begin{equation}
F(\phi)(A''+4 A'^2)+\frac{1}{3}(\phi')^{2}F''(\phi)+\frac{7}{3}
\phi'A'F'(\phi)+\frac{1}{3} \phi''F'(\phi)+\frac{1}{3}
V(\phi)-\frac{1}{6}\sigma(\phi) \delta(z)=0, \label{Ricciii}
\end{equation}
\begin{equation}
F(\phi)(4A''+4A'^2)+\frac{1}{2}(\phi')^{2}
+\frac{4}{3}(\phi')^{2}F''(\phi)+\frac{4}{3}\phi'A'F'(\phi)+\frac{4}{3}\phi''F'(\phi)
+\frac{1}{3}V(\phi)=0, \label{Ricci5}
\end{equation}
and the appropriate boundary conditions on the brane
\cite{KF,Tam}.

Note, that equation (\ref{scalar1}) for the scalar field, is also
satisfied by the background solution, since it is not independent
from the Einstein equations (\ref{Einstein1}); in fact it can be
derived from the latter \cite{KF}.

If one takes into account equations (\ref{Ricciii}) and
(\ref{Ricci5}), the first order Einstein equations (\ref{Efirst}),
can be rewritten in the form:
\begin{eqnarray}
&&2 F(\phi) \left[
-e^{2A}\left(\frac{1}{2}\partial^2_z+2A'\partial_{z}\right)\bar{h}_{ij}
-\frac{1}{2}\Box^{(4)}
\bar{h}_{ij}-\frac{1}{2}\eta_{ij}e^{2A}A'\partial_z(\eta^{kl}\bar{h}_{kl}),
\right. \label{linear1}
\\
 &&-\frac{1}{2}\eta^{kl}(\partial_i\partial_j
\bar{h}_{kl}-\partial_i\partial_k
\bar{h}_{jl}-\partial_j\partial_k \bar{h}_{il})\left.
\right]=\frac{1}{3}\eta_{ij}e^{2A}F'(\phi)\phi'\partial_{z}(\eta^{kl}\bar{h}_{kl})+e^{2A}F'(\phi)\phi'\partial_{z}\bar{h}_{ij}\nonumber
\end{eqnarray}
\begin{eqnarray}
-F(\phi)(\partial^2_z+2A'\partial_{z})\eta^{kl}\bar{h}_{kl}=\frac{1}{3}e^{2A}F'(\phi)\phi'\partial_{z}(\eta^{kl}\bar{h}_{kl}),
\quad \frac{1}{2}\eta^{kl}\partial_z(\partial_k
\bar{h}_{il}-\partial_i \bar{h}_{kl})=0.
\end{eqnarray}

We can simplify the above equations if we perform a gauge
transformation $\bar{h}_{ij}\rightarrow h_{ij}$ (see equations
(\ref{point}), (\ref{gauss0}) and (\ref{gauss1}) below), where the
metric fluctuations $h_{ij}$ in the new coordinate system
satisfies the conditions $\eta_{ij}h^{ij}=0$ (traceless) and
$\partial^{i}h_{ij}=0$ (transverse). In this coordinate system the
Einstein equations decouple and we get:
\begin{equation}
 F(\phi)(-e^{2A}(\partial_{z}^{2}+4
A'\partial_{z})-\Box^{(4)})h_{ij}=e^{2A}F'(\phi)\phi'\partial_{z}h_{ij},
\end{equation}
or equivalently:
\begin{equation}
(\partial_{z}^{2}+ Q'(z)\partial_{z}+e^{-2A}\Box^{(4)})h_{ij}=0,
\label{Sturm1}
\end{equation}
where we have set
\begin{equation}
Q(z) \equiv 4A(z)+ln(F(\phi(z))). \label{def}
\end{equation}
In this paper we assume that the solutions considered satisfy
$F(\phi)>0.$

Note that in this work we do not examine the case of non-zero
scalar field fluctuations, that is we have assumed that the
fluctuation $\tilde{\phi}$ around the background scalar field
vanishes. This is consistent, since one finds that the linearized
equation of motion for the scalar fluctuation reads:
$$\tilde{\phi}^\pp+4 A^\pr \tilde{\phi}^\pr
+\frac{1}{2} \phi^\pr \eta^{ij} h^\pr_{ij}+F^\pr(\phi) R^{(1)}+
F^\pp(\phi) R^{(0)} \tilde{\phi} -V^\pp(\phi) \tilde{\phi} = 0.$$
If we impose the constraint $\tilde{\phi}=0,$ this equation
becomes $\frac{1}{2} \phi^\pr \eta^{ij}
\bar{h}^\pr_{ij}+F^\pr(\phi) R^{(1)} = 0,$ or \be \frac{1}{2}
\phi^\pr \eta^{ij} \bar{h}^\pr_{ij} + \frac{d F(\phi)}{d \phi}
\left[\eta^{ij} \bar{h}^\pp_{ij}+ 5 A^\pr \eta^{ij}
\bar{h}^\pr_{ij} - e^{-2 A} \Box (\eta^{ij} \bar{h}_{ij})- e^{-2
A} \p^i \p^j \bar{h}_{ij} \right] = 0.\label{phi=0} \ee The primes
denote differentiation with respect to $z.$ In the following we
work in a gauge, in which the gravitational perturbation satisfies
the constraints $\eta^{ij} h_{ij}=0, \ \p^i h_{ij}=0$ (see
equation (\ref{gauge}) below). It is easily seen that in this
gauge equation (\ref{phi=0}) is manifestly satisfied.

\section{Graviton localization}

In this section we study the spectrum of metric fluctuations for
brane models with a non-minimally coupled scalar field. We obtain
that the spectrum does include a zero mode localized on the brane
(which is necessary, in order to obtain the 4D Newton's law),
along with a continuum of positive energy states. It should be
noted that in models with a non-minimal coupling the weight
function $r(z)=F(\phi) e^{2A}$ depends on the background scalar
field, which marks an important difference from the RS2-model, or
models with a minimally coupled scalar field.

\subsection{Sturm-Liouville form}

It is easy to find the weight function by bringing the equation
for the gravitational fluctuations in a Sturm-Liouville form. If
we set
\begin{equation}
h_{ij}(x,z)=e^{ipx}u(m,z)
\end{equation}
in equation (\ref{Sturm1}) we obtain
\begin{equation}
(\partial_{z}^{2}+ Q'\partial_{z}+m^{2}e^{-2A})u(m,z)=0,
\label{sturm2}
\end{equation}
where $m^{2}=-p^{i}p_{i}$ is the four-dimensional mass.

We note the following boundary condition on the brane:
\begin{equation}
\partial_{z}h_{ij}(x,z)|_{z=0}=0, \quad {\rm or} \quad u'(m,0)=0.
\end{equation}

Multiplying equation (\ref{sturm2}), with the factor $e^{Q},$
where $Q$ is defined in equation (\ref{def}), we get:
\begin{equation}
-\partial_{z}\left(e^{
Q}\partial_{z}u(m,z)\right)=m^{2}e^{Q-2A}u(m,z).
\end{equation}

The above equation is of the Sturm-Liouville form:
\begin{equation}
-\partial_{z}\left(p(z)\partial_{z}u(\lambda,z)\right)+q(z)
u(\lambda,z)=\lambda r(z) u(\lambda,z), \label{sturm3}
\end{equation}
with the coefficients:
\begin{equation}
p(z)=F(\phi)e^{4A}, \quad r(z)=F(\phi)e^{2A}, \quad q(z)=0, \quad
\lambda=m^{2}.
\end{equation}

The eigenvalue equation (\ref{sturm3}) has a constant solution
$u(0,z)$ for the zero mode $(m^{2}=0).$ This solution $u(0,z)$
must be square integrable with a weight function
$r(z)=F(\phi)e^{2A},$ so the normalizable zero mode is
\begin{equation}
u(0,z)=\frac{1}{\sqrt{\int_{-\infty}^{+\infty}dz F(\phi)e^{2A}}}.
\label{zrmode}
\end{equation}
In addition, the quantity $[u(0,z)]^2 F(\phi)e^{2A} dz$ may be
interpreted as the probability to obtain the corresponding
graviton between the positions $z$ and $z+dz$. Note that we have
considered only solutions which satisfy the restriction
$F(\phi)>0$ and thus the probability density is positive. We have
also assumed that the integral $\int_{-\infty}^{+\infty}F(\phi)
e^{2 A(z)} dz$ exists, that is $F(\phi) e^{2 A(z)}$ tends to zero
quite fast for large $z.$ From the above we conclude that the
eigenvalue equation (\ref{sturm2}) has a \textit{localized }zero
mode which plays the role of the 4D graviton on the brane. As we
show in the next section the remaining spectrum consists of
continuum positive energy states.

\subsection{Schr\"{o}dinger form}

It is instructive to use an alternative form of the equation for
the gravitational fluctuations, called the Schr\"{o}dinger form.
It involves a potential and it offers intuition about the
spectrum.

If one performs the transformation $w=w(z),$ where
$w'(z)=e^{-A(z)},$ equation (\ref{background}) is put into the
conformally flat form:
\begin{equation}
ds^2=e^{2 \tilde{A}(w)}(\eta_{ij}dx^{i}dx^{j}+dw^2), \quad
\tilde{\phi}=\tilde{\phi}(w),
\end{equation}
and equation (\ref{sturm2}) yields:
\begin{equation}
(\partial_{w}^{2}+
\tilde{Q}'(w)\partial_{w}+m^{2})\tilde{u}(m,w)=0, \label{cflat}
\end{equation}
where
\begin{equation}
\tilde{Q}(w)=3\tilde{A}(w)+ln(F(\tilde{\phi}(w))). \label{q1}
\end{equation}
The transformation
\begin{equation}
\tilde{u}(m,w)=e^{-\frac{\tilde{Q}}{2}}\tilde{v}(m,w)
\end{equation}
brings equation (\ref{cflat}) into the Schr\"{o}dinger form:
\begin{equation}
-\partial_{w}^{2}\tilde{v}(m,w)+
\left(\tilde{V}(w)-m^{2}\right)\tilde{v}(m,w)=0 \label{schrform}
\end{equation}
with the potential:
\begin{equation}
\tilde{V}(w)=\frac{1}{2}\tilde{Q}''(w)+\frac{1}{4}(\tilde{Q}'(w))^2.
\label{pot1}
\end{equation}
Equation (\ref{schrform}) can be written alternatively as:
\begin{equation}
L^{\dagger}L\tilde{v}(m,w)=m^2 \tilde{v}(m,w),
\end{equation}
where the operators $L$ and $L^{\dagger}$ are defined through:
\begin{equation}
L^{\dagger} \equiv -\partial_{w}-\frac{1}{2}Q'(w), \quad L \equiv
\partial_{w}-\frac{1}{2}Q'(w).
\end{equation}
As the operator $L^{\dagger}L$ is hermitian and positive definite,
it will have a complete system of eigenstates with non-negative
eigenvalues, or $m^2\geq 0.$ In addition there is a {\em
normalizable} zero mode, which obeys the equation
\begin{equation}
L\tilde{v}(0,w)=0 \quad \Leftrightarrow \quad \tilde{v}(0,w)\sim
e^{\frac{Q(w)}{2}}=e^{\frac{3}{2}\tilde{A}(w)}F(\tilde{\phi}(w)).
\end{equation}

It is readily seen from equations (\ref{q1}) and (\ref{pot1}) that
the potential $\tilde{V}(w)$ strongly depends on
$F(\tilde{\phi}(w)).$

If the potential in equation (\ref{pot1}) vanishes for
$|w|\rightarrow +\infty,$ there exists a continuum spectrum of
positive energy states starting from zero; for extensive
discussion on this point see \cite{Csaki1}. We have checked that
the analytical solutions in \cite{Tam}, with the RS warp factor,
give rise to a potential of the "volcano" form, in agreement with
the above assertions about the spectrum.

\section{Gravity in the RS brane world with a non-minimally coupled bulk scalar field}

In general, brane world models succeed in reproducing the
Newtonian potential on the brane, as they exhibit a localized zero
energy state which mimics the 4-dimensional graviton. In addition,
due to Kaluza-Klein-like excitations, modifications of Newton's
law are predicted at distances smaller than the length scale of
the model.

We would like to examine whether brane models with a non-minimally
coupled bulk scalar field reproduce the 4D gravity on the brane.
The proper approach for the derivation of Newton law has been
presented in \cite{GT} and it is known as the bent brane
formalism; this approach is adopted in the sequel. However, the
bent brane formalism cannot be applied to metrics which are not of
the Randall-Sundrum type in a straightforward way. For this reason
in this paper we examine only single brane solutions with a warp
factor $e^{A(z)}$ of the Randall-Sundrum form $A(z)=-|z|/l$, where
$l$ is the length scale of the model. As already mentioned,
analytical solutions with a Randall-Sundrum type warp factor have
been obtained in \cite{Tam}.

\subsection{Bent brane formalism}

In the previous section we considered a Gaussian normal coordinate
system $(\bar{x}^{i},\bar{z}),$ which is defined by the
hyper-surface $\bar{z}=0$, and we assumed that the brane is
exactly located on the hyper-surface $\bar{z}=0$. In order to
decouple the linearized equations of motion for the metric
fluctuations, we went to another Gaussian normal coordinate system
$(x^{i},z)$, where:
\begin{equation}
x^{i}=\bar{x}^{i}+\xi^{i}, \quad z=\bar{z}+\xi^{5}. \label{point}
\end{equation}
In these new coordinates the gauge conditions:
\begin{equation}
 \eta_{ij}h^{ij}=0 , \qquad  {\rm and} \qquad \partial^{i}h_{ij}=0 \label{gauge}
\end{equation}
are satisfied. In the case of a source-free brane and bulk, the
hyper-surface which defines the new Gaussian normal coordinate
system is also described by the equation $z=0$. However, in the
presence of a point mass source on the brane, with energy momentum
tensor:
\begin{equation}
T^{brane}_{\mu\nu}=S_{\mu\nu}(\bar{x}) \delta(\bar{z}), \quad
S_{\mu\nu}(\bar{x})=
M\delta_{\mu}^{0}\delta_{\nu}^{0}\delta(\bar{\textbf{x}}),
\label{energy1}
\end{equation}
the hyper-surface which defines the new Gaussian normal coordinate
system appears to be bent, and it is described by the equation
$z=-\hat{\xi}^{5}(x)$. We emphasize that the choice
$\hat{\xi}^{5}(x)=0,$ when a mass source is present on the brane,
is not compatible with the gauge conditions of equation
(\ref{gauge}), so that the bending of the brane is unavoidable in
the new coordinate system.

The most general transformations between these two coordinate
systems, which obey the conditions
$\bar{h}_{\mu5}=\bar{h}_{55}=h_{\mu5}=h_{55}=0,$ are:
\begin{equation}
\xi^{i}=-\eta^{ij}\int d\bar{z}e^{-2A(\bar{z})}
\partial_{j}\hat{\xi}^{5}(\bar{x})+\hat{\xi}^{j}(\bar{x}),\quad
\xi^{5}=\hat{\xi}^{5}(\bar{x}), \label{gauss0}
\end{equation}
where the functions $\hat{\xi}^{i}$ and $\hat{\xi}^{5}$ are
independent from the bulk coordinate $\bar{z}$.

The metric fluctuations in the new coordinate system $h_{ij}$
(around the background metric $(\eta_{ij}e^{2 A(z)},1)$), and the
metric fluctuations in the old coordinate system $\bar{h}_{ij}$
(around the metric $(\eta_{ij}e^{2 A(\bar{z})},1)$), are related via
the equation
\begin{equation}
h_{ij}=\bar{h}_{ij}-\partial_{i}\hat{\xi}_{j}-\partial_{j}\hat{\xi}_{i}+2\int
d\bar{z}e^{2A(\bar{z})}\partial_{i}\partial_{j}\hat{\xi}_{5}-2\eta_{ij}A'(\bar{z})
\hat{\xi}^{5} \label{gauss1}
\end{equation}

In the coordinate system, where the position of the brane is at
$\bar{z}=0,$ the junction condition reads:
\begin{equation}
2F(\phi(0))\partial_{\bar{z}}\bar{h}_{ij}|_{\bar{z}=0+}
=-\tilde{S}_{ij}(\bar{x}) \quad \left(\tilde{S}_{\mu\nu} \equiv
S_{\mu\nu}-\frac{1}{3} g_{\mu\nu} S, \quad S \equiv
S^\mu_\mu\right). \label{junction1}
\end{equation}
The above equation is obtained from the linearized Einstein equation
(\ref{linear1}) if we include the source term of Eq.
(\ref{energy1}).

The junction condition for the RS metric in the new coordinate
system obtains by combining equations (\ref{gauss1}) and
(\ref{junction1}):
\begin{equation}
2 F(\phi(0))\partial_{z}h_{ij}|_{z=0+}=-\tilde{S}_{ij}(x)+4
F(\phi(0))\partial_{i}\partial_{j}\hat{\xi}_{5} \label{junction2}
\end{equation}

If we had an arbitrary warp factor $\tilde{A}(\bar{z})$, rather
than the RS-type warp factor $A(\bar{z})=-|\bar{z}|/l,$ an extra
term of the form $-4 F(\phi(0)) \eta_{ij} \tilde{A}''(0^+)
\hat{\xi}^{5} $ would appear in equation (\ref{junction2}), as in
general the second derivative of $\tilde{A}(\bar{z})$ would not be
zero. However, in this case it would be impossible to satisfy
simultaneously the "transverse and traceless" conditions of
equation (\ref{gauge}), and as a result the bent brane formalism
can not be applied in this case in a straightforward way.

Thus, in the new coordinate system, the linearized equation of
motion for the metric fluctuations $h_{ij}:$
\begin{equation}
e^{2A}F(\phi)\partial_{z}^{2}h_{ij}+e^{2A}(4
A'F(\phi)+F'(\phi)\phi')\partial_{z}h_{ij}+F(\phi)\Box^{(4)}h_{ij}=-\Sigma_{ij}(x)\delta(z)
\end{equation}
may be rewritten in the form:
\begin{equation}
F(\phi)\left(\partial_{z}^{2}+Q'\partial_{z}
+e^{-2A}\Box^{(4)}\right)h_{ij}=-\Sigma_{ij}(x)\delta(z),
\label{tensor5}
\end{equation}
where
\begin{equation}
\Sigma_{ij}(x)=\tilde{S}_{ij}(x)-4
F(\phi(0))\partial_{i}\partial_{j}\hat{\xi}_{5}.
\end{equation}
If one defines the retarded five-dimensional Green function
through the equation:
\begin{equation}
F(\phi)\left(\partial_{z}^{2}+Q'\partial_{z}+e^{-2A}\Box^{(4)}\right){\cal
G}^{(R)}_{5}(x,z;x',z')=\delta^{(4)}(x-x')\delta(z-z'),
\end{equation}
the solution of equation (\ref{tensor5}) can be expressed as
\begin{equation}
h_{ij}(x,z)=-\int d^{4}x'{\cal G}^{(R)}_{5}(x,z;x',0)\Sigma_{ij}(x')
\end{equation}
The Green function can be expressed in terms of the complete set
of eigenstates $e^{ipx} u(m,z):$
\begin{equation}
{\cal G}_{5}^{(R)}(x,z;x',z')=-\int\frac{d^4 p
}{(2\pi)^4}e^{ip(x-x')}\left [
\frac{u(0,z)u(0,z')}{p^2}+\sum_{m>0} \frac{u(m,z)u(m,z')}{p^2+m^2}
\right],\label{Green2}
\end{equation}
where $u(m,z)$ satisfies the eigenvalue equation (\ref{sturm2}).
These eigenfunctions should be normalized according to the
equation:
\begin{equation}
\int_{-\infty}^{+\infty}dz \: u(m,z)^2 F(\phi)e^{2A}=1.
\end{equation}
In order to precisely define the summation over states in equation
(\ref{Green2}), it is necessary to consider a regulator brane in
finite proper distance $L$, and then send $L$ to infinity.

Since we will concentrate on static solutions, it is convenient to
define the five-dimensional Green function for the Laplacian
operator:
\begin{equation}
{\cal
G}_{5}(\textbf{x},z;\textbf{x}',z')=\int_{-\infty}^{+\infty}dt'{\cal
G}_{5}^{(R)}(x,z;x',z').
\end{equation}
If we perform the integration over $p$ in equation (\ref{Green2})
and set $r=|\textbf{x}-\textbf{x}'|$, we obtain
\begin{equation}
{\cal G}_{5}(\textbf{x},z;\textbf{x}',z')=-\frac{1}{4\pi
r}\left[u(0,z)u(0,z')+ \;\sum_{m>0}  u(m,z)u(m,z') e^{-m r}
\right]. \label{Green1}
\end{equation}
The metric fluctuation $\bar{h}_{ij}$ on the brane at $\bar{z}=0$
can be obtained from equation (\ref{gauss1}):
\begin{equation}
\bar{h}_{ij}(x,0)=h_{ij}(x,0)+\partial_{i}\hat{\xi}_{j}+\partial_{j}\hat{\xi}_{i}-2\left[\int
d\bar{z}e^{2A(\bar{z})}\right]_{\bar{z}=0}\partial_{i}\partial_{j}\hat{\xi}_{5}+2\eta_{ij}A'(0)
\hat{\xi}^{5}. \label{tensor1}
\end{equation}
If we use the remaining gauge freedom and choose
$\hat{\xi}^{i}(x)$ according to
\begin{equation}
\hat{\xi}_{i}=\partial_{i}\left(\left[\int
d\bar{z}e^{2A(\bar{z})}\right]_{\bar{z}=0}\hat{\xi}_{5}-2
F(\phi(0))\int d^{3}\textbf{x}'{\cal
G}_{5}(\textbf{x},0;\textbf{x}',0) \hat{\xi}_{5}\right)
\end{equation}
(see \cite{Csaki}), we obtain a simple expression for the
fluctuation $\bar{h}_{ij}(x,0)$ on the brane:
\begin{equation}
\bar{h}_{ij}=-\int d^{3}\textbf{x}'{\cal
G}_{5}(\textbf{x},0;\textbf{x}',0)\tilde{S}_{ij}(\textbf{x}')+2\eta_{ij}A'(0)
\hat{\xi}^{5}. \label{tensor2}
\end{equation}
The function $\hat{\xi}^{5}$ can be determined if take into
account that in the new coordinate system the condition
$h_{i}^{i}=0$ must be satisfied. Then from equation
(\ref{junction2}) for $h_{ij}$ we obtain that $\hat{\xi}^{5}$ is a
solution of the equation
\begin{equation}
\Box^{(4)} \hat{\xi}^{5}=-\frac{1}{12} F(\phi(0))^{-1}S,
\label{bend1}
\end{equation}
where $S=S_i^i$.
It is not difficult to show that the condition
$\partial^{i}h_{ij}=0$ is also satisfied.

In the case of a point mass source on the brane, with the energy
momentum tensor given in equation (\ref{energy1}), we obtain:
\begin{equation}
\hat{\xi}^{5}=-\frac{M}{48 \pi F(\phi(0))}\frac{1}{r}.
\label{bending1}
\end{equation}
Using equations (\ref{Green1}), (\ref{tensor2}) and
(\ref{bending1}) we get
\begin{equation}
\bar{h}_{00}=\frac{M u(0,0)^2}{6 \pi r}+\frac{M} {6 \pi
r}\sum_{m>0} u(m,0)^2 e^{-m r}-\frac{M}{24 \pi l
F(\phi(0))}\frac{1}{r}. \label{tensor3}
\end{equation}
Notice that the Newton potential is given by: $$V(r) =
\frac{\bar{h}_{00}}{2}.$$ Taking into account that the
four-dimensional Newton's constant $G_{4}$ is defined by the
dimensional reduction equation \cite{Mav}
\begin{equation}
\int dz d^{4}x \;\sqrt{|g|}F(\phi)R+...=\int_{-\infty}^{+\infty}
dz \; F(\phi) e^{2 A} \int d^{4}x \sqrt{|g^{(4)}|}R^{(4)}+...
\equiv \frac{1}{16 \pi G_{4}}\int d^{4}x
\sqrt{|g^{(4)}|}R^{(4)}+...
\end{equation}
we obtain:
\begin{equation}
\frac{1}{16 \pi G_{4}}=\int_{-\infty}^{+\infty}dz \;e^{2
A}F(\phi)=\frac{1}{u(0,0)^{2}}, \label{reduction}
\end{equation}
where use has been made of equation (\ref{zrmode}).

Equation (\ref{tensor3}) gives:
\begin{equation}
\bar{h}_{00}=\frac{2 M G_{4} (1+\alpha_{F})}{r}+\frac{8M
G_{4}}{3r}\sum_{m>0} \frac{u(m,0)^2}{u(0,0)^2}e^{-m r},
\label{tensor4}
\end{equation}
where
\begin{equation}
\alpha_{F}=\frac{1}{3}\left(1-\frac{\int_{-\infty}^{+\infty} dz\:
e^{2A}F(\phi)}{l F(\phi(0))}\right)=\frac{\xi k
\phi(0)^2\left(-1+\int_{-\infty}^{+\infty} d\hat{z}\:
e^{2A}\frac{\phi(\hat{z})^2}{\phi(0)^2}\right)}{3(1-\xi k
\phi(0)^2)}, \quad {\rm for} \ \phi(0)\neq 0, \label{cor1}
\end{equation}
\begin{equation}
\alpha_{F}=\frac{\xi k}{3} \int_{-\infty}^{+\infty} d\hat{z}\:
e^{2A} \phi(\hat{z})^2, \quad {\rm for} \ \phi(0)=0. \label{cor2}
\end{equation}
We have used the notation: $\hat{z} \equiv \frac{z}{l}.$ We point
out that the first term in equation (\ref{tensor4}), namely the
Newtonian potential, arises as a combination of two contributions,
the first one is that of the zero mode, while the second one is
due to the brane bending term of equation (\ref{bending1}).

The second term, on the other hand, involving the tower of the
massive states, gives rise to corrections to Newton's law. We can
use a second regulator brane in order to express the summation
over states as an integral. Now, use of dimensional analysis
indicates that this term tends to zero for $r>>l$ (where $l$ is
the $AdS_5$ radius), while for small $r$ ($r<<l$) the corrections
become very important; for example, for $r<<l$ they should modify
the potential to its five-dimensional version, namely
$\frac{1}{r^2}.$ However, precise knowledge of the corrections
presupposes knowledge of the eigenfunctions. This can be done
analytically for the RS2-model, but in our case only numerical
calculations are possible.

We would like to emphasize that in this paper we will severely
narrow the acceptable models; as a result the corrections of
Newton's law are expected to be quite similar to those of the
RS2-model, which have already been found analytically \cite{CR}. For
this reason we will not attempt a numerical computation of the
summation over the massive states.

\subsection{Zero mode truncation}

For large distances the dominant part of the five dimensional
Green function, for $|\textbf{x}-\textbf{x}'|\gg l$, is due to the
contribution of the zero mode part, as the remaining part is
suppressed by an $O(l)$ factor. Thus, if we neglect the
contribution of the continuous modes, which is of the order of
$O(l)$, we obtain
\begin{equation}
{\cal G}_{5}(\textbf{x},0;\textbf{x}',0)=u(0,0)^{2}{\cal
G}_{4}(\textbf{x},\textbf{x}'), \quad  \Box^{(4)}_{\textbf{x}} {\cal
G}_{4}(\textbf{x},\textbf{x}')=
\delta^{(4)}(\textbf{x}-\textbf{x}')\label{red}
\end{equation}

From equations (\ref{red}), (\ref{tensor2}) and (\ref{bend1}) we
derive the equation that should be obeyed by the gravitational
perturbation $\bar{h}_{ij}:$
\begin{equation}
\Box^{(4)}\bar{h}_{ij}=-16 \pi G_{4}(S_{ij}-\frac{1}{2}\eta_{ij}S)-8
\pi \;  G_{4} \; \alpha_{F} \;\eta_{ij}S \label{mod1}
\end{equation}
We observe that the above equation is somewhat different from the
linearized equation of standard 4D general relativity:
\begin{equation}
\Box^{(4)}\bar{h}_{ij}=-16 \pi G_{4}(S_{ij}-\frac{1}{2}\eta_{ij}S) \label{nomod}
\end{equation}
Note that this situation, of modified linearized equations on the
brane, is quite similar with the case of RS1-model; see \cite{GT}.

For a point-like source, the solutions of Eq. (\ref{mod1}) read:
\begin{equation}
\bar{h}_{00}=\frac{2(1+\alpha_{F})M G_{4}}{r}, \quad
\bar{h}_{ij}=\frac{2(1-\alpha_{F})M G_{4}}{r}\eta_{ij}.
\end{equation}
Absorbing the factor $(1+\alpha_{F})$ into the mass $M$ (see
\cite{sw}), the solution can take the form:
\begin{equation}
\bar{h}_{00}=\frac{2MG_{4}}{r}, \quad
\bar{h}_{ij}=\gamma_{F}\frac{2MG_{4}}{r}\eta_{ij}
\end{equation}
where
\begin{equation}
\gamma_{F}=\frac{1-\alpha_{F}}{1+\alpha_{F}} \label{aF}
\end{equation}

We recall that the standard isotropic form of the metric can be
expanded to yield the expressions:
\begin{equation} \bar{h}_{00}=\left(-1+\frac{2 M G_4}{r} - \beta
\frac{2 M^2 G_4^2}{r^2}+\dots\right),\; \bar{h}_{ij} =\left(1 +
\gamma \frac{2 M G_4}{r} + \dots\right)\eta_{ij},   \ i,j=1,2,3.
\label{expBD}
\end{equation}
giving a fairly general version of the metric \cite{sw}. Recalling
the Schwarzschild solution, one may check that the standard
Einstein equations predict $\beta=1, \ \gamma=1.$

The expansion (\ref{expBD}) can also represent alternative gravity
theories. For instance, the Brans-Dicke theory would yield
$\alpha=1, \ \beta=1, \ \gamma =
\frac{\omega_{BD}+1}{\omega_{BD}+2},$ where $\omega_{BD}$ is the
Brans-Dicke parameter. In our case we have not yet calculated the
parameter $\beta,$ since that would require second order
perturbation theory \cite{sec}  and is deferred to a future publication. If we
assume that $\beta=1,$ our result (\ref{aF}) is compatible with a
four-dimensional Brans-Dicke theory, hence we obtain that the
Brans-Dicke parameter is
\begin{equation}
 \omega_{BD}=\frac{1}{2 \alpha_{F}}-\frac{3}{2} \label{bd1}
\end{equation}

A lower bound on $\omega_{BD},$ necessary for consistency with
solar system measurements, which appears in the literature
\cite{R2} reads: $\omega_{BD}\geq 500.$  Then, with the help of
equation (\ref{bd1}) this restriction becomes:
\begin{equation}
0\leq\alpha_{F}\leq 10^{-3}, \label{res0}
\end{equation}
or equivalently:
\begin{equation}
0\leq\frac{\xi k \phi(0)^2\left(-1+\int_{-\infty}^{+\infty}
d\hat{z}\: e^{2A}\frac{\phi(\hat{z})^2}{\phi(0)^2}\right)}{3(1-\xi
k \phi(0)^2)}\leq 10^{-3}, \quad {\rm for} \ \phi(0)\neq 0,
\label{res1}
\end{equation}
\begin{equation}
0\leq\frac{\xi k}{3} \int_{-\infty}^{+\infty} d\hat{z}\: e^{2A}
\phi(\hat{z})^2 \leq 10^{-3}, \quad {\rm for} \ \phi(0)= 0.
\label{res2}
\end{equation}
We conclude that static solutions of brane models with non-minimal
coupling, with a warp factor of the RS type $A(z)=-|z|/l$ ,
\textit{are acceptable only if they satisfy the conditions}
(\ref{res1}) or (\ref{res2}), otherwise we cannot recover general
relativity on the brane. In other publications \cite{cwill}, the
lower bound for $\omega_{BD}$ is set to even larger numbers. If
these numbers are adopted, the restrictions for our model will
become even stronger.

A complete investigation of the restriction (\ref{res0}) is beyond
the scope of this paper, as the analytical static solutions of the
model depend on a large number of free parameters \cite{Tam}.
However, one may describe two circumstances, in which equation
(\ref{res1}) is satisfied: (a) If
$\left|-1+\int_{-\infty}^{+\infty} d\hat{z}\:
e^{2A}\frac{\phi(\hat{z})^2}{\phi(0)^2}\right| \sim 1$, then it is
necessary that $|\xi| k \phi(0)^2 \sim 10^{-3}.$ (b) If equation
(\ref{res1}) is to be satisfied for $|\xi| k \phi(0)^2$ of order
one, then another type of fine tuning appears, namely one should
have $\left|-1+\int_{-\infty}^{+\infty} d\hat{z}\:
e^{2A}\frac{\phi(\hat{z})^2}{\phi(0)^2}\right| \sim 10^{-3}.$ We
observe that equation (\ref{res1}) is satisfied only if the
quantities $\left(-1+\int_{-\infty}^{+\infty} d\hat{z}\:
e^{2A}\frac{\phi(\hat{z})^2}{\phi(0)^2}\right)$ and $\xi k
\phi(0)^2 $ have the same sign. Numerical study of the analytical
solutions contained in \cite{Tam}, indicates that these models
fall into case (a).

We would like to emphasize that in other publications \cite{cwill},
the lower bound for $\omega_{BD}$ is set to even larger numbers, for
example $\omega_{BD}>10^5$. If these numbers are adopted, the
restrictions for our model will become even stronger, and as a
result the class of acceptable models will be severely narrow, and
quite closely to RS2-model.

\section{Conclusions}

We have studied gravitational perturbations for a category of
brane models involving a scalar field non-minimally coupled with
gravity. The focus of our work has been on the study of the
resulting gravitational potential, in particular its comparison
against the Newtonian one: we considered the effect of a point
mass at rest on the brane and used the bent brane formalism. It
turned out that the perturbation $\bar{h}_{ij}$ belongs to a more
general class of metrics, also including the Brans-Dicke theory.
Observational constraints yield restrictions on the model, which
deserve further investigation to determine the region of the
parameter space which is relevant for model building. These
restrictions constitute a test, which should be passed before any
theory belonging to this category is further considered.

As we mentioned in previous sections, in the case of the bent brane formalism
the gauge is fixed from the beginning. In particular, we have chosen
a specific gauge ($h_{5\mu}=0$), which holds in Gaussian normal coordinates.
The remaining gauge freedom (see Eq. (49)) can be used for
imposing the transverse and traceless conditions, where
the equations of motion decouple. However, one could work in a gauge invariant formalism as is done in Ref. \cite{MG}.
In this case graviphotons, graviscalar
and the perturbation of the bulk scalar field are mixed in complicated equations of motion.
Of course, the gauge choice in our paper is not enough to cover completely this general case. New solutions may exist, 
where the contribution of these additional fields is significant.

\appendix

\section{Appendix: Einstein Frame}

In this appendix we derive the linearized equation for metric
fluctuations by working in the Einstein Frame, rather than the
Jordan frame which has mainly been used in this paper. In
particular, we show that the descriptions of the metric
fluctuations in the two frames are equivalent.

Using the conformal transformation
\begin{equation}
\tilde{g}_{\mu\nu}=\omega^2(x) g_{\mu\nu} \label{conf},
\end{equation}
one may transform the action of Eq. (\ref{actio1}) to a minimally
coupled form. We refer to $g_{\mu\nu}$ as the Jordan frame metric,
while $\tilde{g}_{\mu\nu}$ is the corresponding metric in the
Einstein frame.

In this paper we have studied tensor fluctuations in the Jordan
frame, and we determined the corresponding linearized equation
(\ref{cflat}). However, one might transform equation (\ref{conf})
to the Einstein frame, where the action is reduced to a minimally
coupled form. The derivation of the linearized Einstein equations
has already been done previously (see, for example, \cite{wolfe}).

In the Einstein frame we assume the conformal background metric
with fluctuations of the form:
\begin{equation}
ds_{EF}^2=e^{2
\tilde{B}(w)}\left((\eta_{ij}+h_{ij})dx^{i}dx^{j}+dw^2\right).
\end{equation}
If we set
\begin{equation}
h_{ij}(x,w)=e^{ipx}\tilde{u}(m,w)
\end{equation}
we obtain the linearized equation for the metric
fluctuations (see for example \cite{wolfe},
where a minimally coupled model is examined)
\begin{equation}
(\partial_{w}^{2}+3 \tilde{B}'(w) \partial_{w}+m^{2})\tilde{u}(m,w)=0.
\label{ex1}
\end{equation}
The question that arises is whether our results in the Jordan
frame are compatible with the corresponding results in the
Einstein frame (\ref{ex1}).

If we just consider the gravity part of the action and impose the
condition
\begin{equation}
 S=\int d^{5}x \;\sqrt{|g|}\left(
F(\phi)R+...\right)=\int d^{5}x \;\sqrt{|\tilde{g}|}\left(
\tilde{R}+...\right)
\end{equation}
taking into account that
\begin{equation}
\tilde{g}_{\mu\nu}=\omega^2(x) g_{\mu\nu}, \quad
\tilde{R}=\omega^{-2}(x) R, \quad
\sqrt{|\tilde{g}|}=\omega^{5}(x)\sqrt{|g|} \end{equation} we
obtain:
\begin{equation}
\omega^2(x)=\left(F(\phi)\right)^{2/3}. \label{wm}
\end{equation}
Assuming that the conformal background metric in the Jordan frame
has the form
\begin{equation}
ds_{JF}^2=e^{2 \tilde{A}(w)}\left((\eta_{ij}+h_{ij})dx^{i}dx^{j}+dw^2\right)
\end{equation}
then in the Einstein frame, with the help of equations
(\ref{conf}) and (\ref{wm}), we obtain:
\begin{equation}
ds_{EF}^2=\left(F(\phi)\right)^{2/3} e^{2 \tilde{A}(w)}\left((\eta_{ij}+h_{ij})dx^{i}dx^{j}+dw^2\right)
\end{equation}
or equivalently:
\begin{equation}
\tilde{B}(w)=\tilde{A}(w)+\frac{1}{3}\ln\left(F(\phi)\right)
\end{equation}
If we substitute the above equation in (\ref{ex1}), we obtain
\begin{equation}
(\partial_{w}^{2}+
\tilde{Q}'(w)\partial_{w}+m^{2})\tilde{u}(m,w)=0,
\end{equation}
where
\begin{equation}
\tilde{Q}(w)=3\tilde{A}(w)+ln(F(\tilde{\phi}(w))).
\end{equation}
which coincides with equation (\ref{cflat}), proving that the
descriptions of the tensor fluctuations in the Jordan and the
Einstein frames are equivalent.

\section{Acknowledgements}
This work is supported by the EPEAEK programme "Pythagoras II" and
co-funded by the European-Union (75$^{\circ}/_{\circ}$) and the
Hellenic state (25$^{\circ}/_{\circ}$).  We thank the anonymous
referee for useful suggestions.

\end{document}